\documentclass[prl,aps,twocolumn,groupedaddress,floats,showpacs,final,superscriptaddress]{revtex4-2}
\usepackage[latin3]{inputenc}
\usepackage[makeroom]{cancel}
\usepackage{graphicx}
\usepackage{amsmath}
\usepackage{amsfonts}
\usepackage{amssymb}
\usepackage{chngcntr}
\usepackage{color}
\usepackage[left=2cm,right=2cm,top=2cm,bottom=2cm]{geometry}

\usepackage{graphicx} 
\usepackage{dcolumn}
\usepackage{bm}
\usepackage{simplewick}
\usepackage{array}
\usepackage{appendix}

\RequirePackage[
   hyperindex,colorlinks,bookmarksnumbered,
   plainpages=true,pdfstartview=FitH]{hyperref}
\hypersetup{linkcolor=blue,urlcolor=blue,citecolor=blue}
\usepackage{hyperref}
\usepackage{mathrsfs}
\definecolor{purple}{rgb}{0.5,0,0.6}

\usepackage{ulem}
\renewcommand{\emph}[1]{\textit{#1}}
\definecolor{darkblue}{rgb}{0,0,0.5}
\definecolor{darkgreen}{rgb}{0,0.5,0}
\definecolor{darkred}{rgb}{.7,0,0}
\definecolor{purple}{rgb}{0.5,0,0.6}
\definecolor{orange}{rgb}{1,0.5,0}
\definecolor{grey}{rgb}{.6,.6,.6}
\definecolor{lightpink}{rgb}{1,0.7,0.75}
\definecolor{pink}{rgb}{1,0.4,0.58}
\definecolor{deeppink}{rgb}{1,0.08,0.58}

\usepackage{ulem}

\begin{document}

\date{\today}
\title{Probing Luttinger Liquid Properties in Multichannel Two-Site Charge Kondo Simulator}

\author{A. V. Parafilo}
\email{aparafil@ibs.re.kr}
\affiliation{Center for Theoretical Physics of Complex Systems, Institute for Basic Science, Expo-ro 55, Yuseong-gu, Daejeon 34126, Republic of Korea}

\author{V.~M.~Kovalev}
\affiliation{Novosibirsk State Technical University, Novosibirsk 630073, Russia}

\author{I.~G.~Savenko}
\affiliation{Department of Physics, Guangdong Technion -- Israel Institute of Technology, 241 Daxue Road, Shantou, Guangdong, 515063, China}
\affiliation{Technion -- Israel Institute of Technology, 32000 Haifa, Israel}
\affiliation{Guangdong Provincial Key Laboratory of Materials and Technologies for Energy Conversion, Guangdong Technion--Israel Institute of Technology, Guangdong 515063, China}

\date{\today}

\begin{abstract}
We study the influence of many-body interactions on the transport properties in a two-site charge Kondo circuit recently implemented 
in a hybrid metal-semiconductor double-quantum dot device 
[W.~Pouse {\it et al.}, Nat.~Phys.~{\bf 19}, 492 (2023)]. 
There emerge two principal types of interactions: (i) an intrinsic one, described by the Luttinger liquid model, and (ii) an induced one, which appears due to the coupling of the system to an Ohmic environment. 
Case (i) could be achieved if the charge Kondo circuit operates in the fractional quantum Hall regime, while case (ii) can be implemented via a finite number of open ballistic channels coupled to both the quantum dots. 
We demonstrate that the conductance scaling for the case of strong and weak interdot coupling is fully determined by the effective interaction parameter, which is the combination of the fractional filling factor $\nu=1/m$ and the number of transmitting channels. 
Furthermore, we predict that the fractional filling factor $\nu$ defines a universal Kondo scaling in the vicinity of a special triple quantum critical point featured by the emergence of a $\mathbb{Z}_3$ parafermion.
\end{abstract}

\maketitle

The Kondo model introduced in a pioneering work~\cite{kondo} allowed for the explanation of the minimum in the dependence of resistance on temperature in dilute magnetic alloys. 
Soon, it became clear that this model represents a powerful tool giving access to a wide range of strongly correlated phenomena~\cite{hewson}, providing a testbed for various many-body theoretical methods~\cite{nozieres, anderson, wilson, andrei, wigman, blandin}. The rebirth of the model occurred with the development of mesoscopic physics, when it became clear that a quantum dot (QD), a principal element of the single-electron transistor~\cite{shekhteradd, shekhter2, kastner}, under certain conditions behaves as a quantum spin-1/2 impurity, thus manifesting the properties of the Kondo model~\cite{revival, glazmanraikh, Gordon_kondoexp, kouw, Wiel}. 

It turned out, that the Kondo model has many different implementations: Kondo-like effects occur in systems where few local degrees of freedom are coupled to one or more bath continua~\cite{legget, weiss, lehur_review}. 
One such system, which we focus on in this Letter, is called {\it the charge Kondo} (CK) circuit.
It was proposed by Matveev, Flensberg, and Furusaki in the early 1990s~\cite{matveev1, flensberg, matveev2, furusakimatveev}. 
They considered a large QD formed in a two-dimensional electron gas (2DEG) connected to the leads via single-mode quantum point contacts (QPCs). 
In contrast to the original Kondo problem, in which the Kondo effect is attributed to the spin degree of freedom, a quantum pseudo-spin in the charge Kondo implementation is represented by two degenerate macroscopic charge states of the QD~\cite{matveev1,matveev2}. 

The first experimental implementation of the CK device represented a hybrid metal-semiconductor single-electron transistor on the base of a GaAs/AlGaAs heterostructure operating in the integer quantum Hall (IQH) regime~\cite{pierre1, pierre2, pierre3}. 
Due to remarkable control over the conducting channels entering the QD, the CK device in~\cite{pierre1, pierre2, pierre3} provides unprecedented access to multichannel Kondo physics. 
A breakthrough series of  experiments devoted to various aspects of a two- and three-channel CK problem \cite{pierre1, pierre2, pierre3, pierre4, pierre5} induced enormous interest among  the theoretical~\cite{thanh2018, thanhprl,VN, parafilo, parafilo2, parafilo3, parafilo4, sim, sim2, sela1, sela2, fritz, kyrylo, karki1, karki2, karki3, kiselev, florian} and experimental~\cite{heatcoulomb, pierre, inter, DCB, DGG} scientific community in the field. 

A very recent experiment~\cite{DGG} suggested a two-site (two-QD) generalization of the CK circuit as a platform for probing frustrated interaction at the exotic quantum critical point with fractional excitation~\cite{karki1,  karki2}. 
The circuit represents two hybrid metal-semiconductor islands coupled in series via QPCs, with each island connected to its own lead.
In a way, such a setup `simulates' the competition between the Ruderman-Kittel-Kasuya-Yosida (RKKY) interaction of two QDs/impurities and an individual Kondo screening of each QD/impurity~\cite{jones,twoimp}. 
It should be noted that a two-site CK (2SCK) circuit~\cite{DGG} with weak and strong interdot couplings was initially proposed in Ref.~\onlinecite{thanh2018} to identify the competition between Fermi- and non-Fermi liquid behavior 
featured by individual 
multichannel CK simulators. 

In this Letter, we address a fundamental question: How does the many-body interaction in the conducting channels of the circuit~\cite{DGG} affect low-temperature transport properties in the 2SCK device? (Is it possible to probe the competition between Fermi- and non-Fermi liquid behavior of a 2SCK device via conductance scaling?) Note that the influence of many-body interactions on the properties of the single-site CK circuit was studied in a series of works \cite{VN,parafilo,parafilo2,parafilo3,parafilo4}.
To answer the question for the case of 2SCK circuit, we examine two types of interaction: (i) an 'intrinsic' one, which we describe by the Luttinger liquid (LL) model~\cite{Tomonaga, Luttinger, Haldane, gogolin, giamarchi}, and (ii) an `induced' one, which may appear due to the coupling of the coherent conductors to an Ohmic dissipative environment~\cite{weiss, devoret, safisaleur, pierre, DCB}. 
The first type of interaction can be experimentally studied in the setup~\cite{DGG} in the fractional quantum Hall (FQH) regime~\cite{chang}. 
The second type of interaction can be implemented by controlling the number of open ballistic channels connected to the QDs. 
We show, that the interaction mechanisms mentioned above uniquely determine the quantum transport behavior of the system, in particular, the maximum of the conductance and the temperature scaling. 

\begin{figure}
\centering 
\includegraphics[width=1\columnwidth]{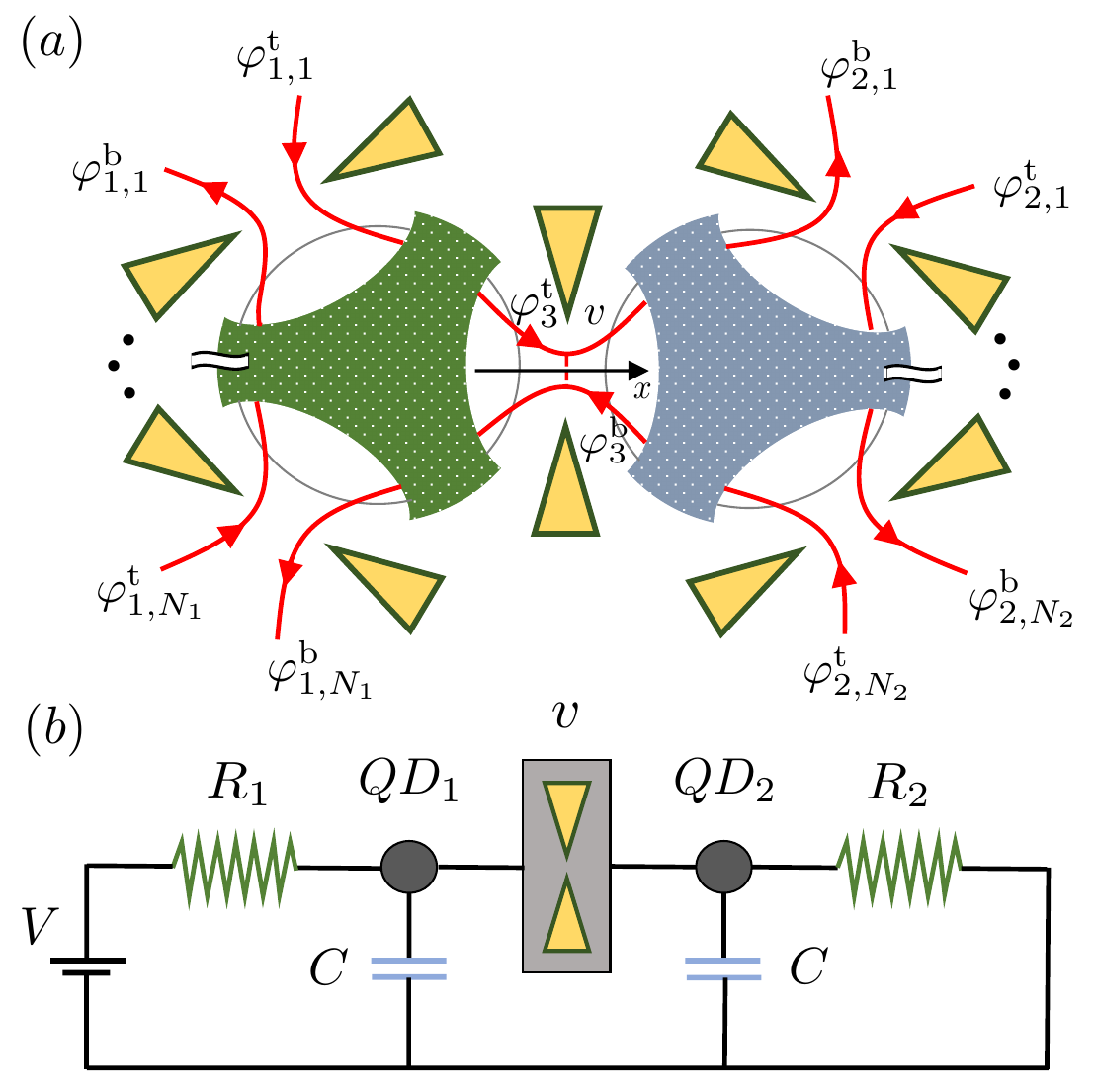} \caption {(a) System schematic: a hybrid metal-semiconductor device 
consists of two metallic quantum dots (QDs) strongly coupled via a single-mode quantum point contact (QPC). 
The $2N_1$ and $2N_2$ edge modes ballistically enter the first and the second QD, respectively. 
The system is exposed to a strong magnetic field, which drives it in the fractional quantum Hall regime. 
In this regime, 1D chiral edge (Laughlin) excitations with a fractional filling factor $\nu=1/m$ (with $m$ an odd integer) flow along the top and the bottom edges.
They are characterized by the bosonic fields $\varphi_{\alpha,i}^{\textrm{t}(\textrm{b})}(x)$.   
(b) An equivalent electric circuit of the device: a conducting channel with a backscattering coefficient $v$ formed in the Luttinger liquid with the interaction constant $\nu$ is coupled to two resistors $R_1=(\nu N_1 G_0 )^{-1}$ and $R_2=(\nu N_2 G_0 )^{-1}$ (with $G_0=R_0^{-1}=e^2/2\pi\hbar$). 
This circuit corresponds to the Luttinger liquid with the weak potential barrier and an effective interaction parameter $K=1/(\nu^{-1}+r_1+r_2)$, where $r_i=R_i/R_0$~\cite{safisaleur}.  }
\label{Fig1} 
\end{figure}

\textit{ Model.---} 
Consider a multichannel 2SCK circuit -- a hybrid metal-semiconductor device consisting of two large metallic islands or 
QDs incorporated into the 2DEG hosted in a GaAs/AlGaAs heterostructure [Fig.~\ref{Fig1}(a)]. 
The circuit is in the FQH regime since it is exposed to a strong magnetic field. 
Furthermore, we assume that the two QDs are strongly coupled via a single-mode QPC, while $2 N_1$ and $2N_2$ edge modes freely enter the first and the second QDs, respectively.
Suppose an external magnetic field and the density of 2DEG satisfy such conditions that the FQH has a filling factor $\nu=1/m$, where $m$ is an odd integer. 
In this regime, 1D chiral Laughlin \cite{laughlin} excitations conduct the fractional charge $e^{\ast}=\nu e$ along the top and bottom edge channels. 

The low-energy physics of the FQH edge states can be characterized by the scalar bosonic fields $\varphi^{\rm t(b)}_{\alpha,i}(x)$ \cite{wenadd,wen}, which satisfy the Kac-Moody commutation relation $[\partial_x\varphi_{\alpha,i}^{\rm t(b)}(x),\varphi_{\alpha',i'}^{\rm t(b)}(x')]=\pm 2i\pi\nu\delta_{\alpha\alpha'}\delta_{ii'}\delta(x-x')$. 
Here, $\varphi^{\rm t(b)}_{1,i}$($\varphi^{\rm t(b)}_{2,i}$) denotes the $N_1$($N_2$) incoming and $N_1$($N_2$) outgoing edge states, which enter the first (second) QD, while $\varphi_{3,i}^{\rm t(b)}$ describe the modes connecting both QDs. The whole system can be described by the Hamiltonian $H=H_0+H_C+H_{BS}$, where
\begin{eqnarray}
H_0=\frac{\hbar v_F}{4\pi \nu}\sum_{\alpha=1}^3\sum_{i=1}^{N_{\alpha}}\int dx \left\{\left(\partial_x \varphi_{\alpha,i}^{\textrm{t}}\right)^2+\left(\partial_x \varphi_{\alpha,i}^{\textrm{b}}\right)^2\right\}
\end{eqnarray}
is the kinetic energy of the top and bottom fractional edge excitation, and $v_F$ is the Fermi velocity. Since $N_3=1$, we use the simplified notation $\varphi_{3,1}^{\rm t(b)}\equiv \varphi_{3}^{\rm t(b)}$ in what follows. 

Two large metallic islands (QDs) can be described by the mesoscopic Coulomb blockade Hamiltonian,
\begin{eqnarray}\label{hamC}
H_C=\frac{1}{2C}\left(Q_1^2+Q_2^2\right),
\end{eqnarray}
where $C$ is the capacitance of metallic islands, and
\begin{eqnarray}
&&Q_1=e \sum_{\sigma={\rm t,b}} \left[\sum_{i=1}^{N_{1}}\int_0^{\infty} dx\, \rho_{1,i}^{\sigma}(x)+\int_{-\infty}^{0} dx\, \rho_{3}^{\sigma}(x)\right],\nonumber\\
&&Q_2=e \sum_{\sigma={\rm t,b}} \left[\int_0^{\infty} dx\, \rho_{3}^{\sigma}(x)+\sum_{i=1}^{N_2}\int_{-\infty}^0 dx\, \rho_{2,i}^{\sigma}(x)\right]\nonumber
\end{eqnarray}
are the charges accumulated at the first and second QD, respectively, while $\rho_{\alpha,i}^{\rm t(b)}=\pm (1/2\pi) \partial_x \varphi_{\alpha,i}^{\rm t(b)}$ is the density operator \cite{footnote3}. 

A weak tunneling of the Laughlin quasiparticles between counter-propagating (top and bottom) channels at the QPC placed between two QDs is covered by the Hamiltonian
\begin{eqnarray}\label{hB}
H_{BS}=\frac{D}{2\pi}v\left[ e^{i \varphi_3^{\textrm{t}}(0)-i\varphi_3^{\textrm{b}}(0)}+\textrm{H.c.}\right],
\end{eqnarray}
where $v\ll1$ is the backscattering matrix element, and $D$ is the energy cutoff.

In the spirit of early works~\cite{flensberg,furusakimatveev,andreevmatveev}, 
it is more common to describe the system via an Euclidean action $\mathcal{S}=\mathcal{S}_0+\mathcal{S}_C+\mathcal{S}_{BS}$. In this case, combining the top and bottom chiral fields $\varphi^{\rm t(b)}_{\alpha,i}(x)$ is more convenient to represent the action in terms of new 
variables: $\phi_{\alpha,i}=(\varphi_{\alpha,i}^{\rm t}-\varphi_{\alpha,i}^{\rm b})/2$ and $\theta_{\alpha,i}=(\varphi_{\alpha,i}^{\rm t}+\varphi_{\alpha,i}^{\rm b})/2\nu$, which obey the canonical commutation relation $[\phi_{\alpha,i}(x),\partial_x \theta_{\alpha',i'}(x')]=i\pi\delta_{\alpha\alpha'}\delta_{ii'}\delta(x-x')$. As a result, the free Euclidean action of the Laughlin edge modes $\mathcal{S}_0$ acquires the form of a {\it non-chiral Luttinger liquid} action \cite{giamarchi} (in $\hbar=k_B=1$ units),
\begin{eqnarray}
\label{freeaction}
\mathcal{S}_{0} =\frac{v_F}{2\pi K}\sum_{\alpha=1}^{3}\sum_{i=1}^{N_{\alpha}} \int dx\int\limits_{0}^{\beta}d\tau \left(\frac{[\partial_{\tau}\phi_{\alpha,i}]^{2}}{v_F^2}+ [\partial_{x}\phi_{\alpha,i}]^{2}
\right),
\end{eqnarray}
where 
the Luttinger interaction parameter $K$$=$$\nu$ and $\beta=T^{-1}$ is the inverse temperature. It should be noted that the fields $\theta_{\alpha,i}(x,\tau)$ were integrated out in the derivation of Eq.~(\ref{freeaction}); see, e.g.,~\cite{giamarchi} and the Supplemental Material~\cite{suplement}. 


The action of the double-QD Coulomb blockade corresponding to Eq.~(\ref{hamC}) has the form
\begin{eqnarray}\label{aC}
&&\mathcal{S}_{C} = \frac{E_{C}}{\pi^2}\int_{0}^{\beta}d\tau\left[\hat{n}^2_1+\hat{n}_2^2\right],\label{coulomb}
\end{eqnarray}
where $E_C=e^2/2C$ is the charging energy of two Ohmic contacts (with zero-energy level spacing) \cite{footnote3},
and
\begin{eqnarray}
\hat{n}_{1(2)}=\pm\left[\phi_{3}(0,\tau)-
\sum_{i=1}^{N_{1(2)}}
\phi_{1(2),i}(0,\tau)\right]\nonumber
\end{eqnarray}
is the total number of quasiparticles that enter the first (second) QD.

Finally, $\mathcal{S}_{BS}$ is the small contribution to the action $\mathcal{S}$ due to backscattering at the QPC,
\begin{eqnarray}\label{backsc}
\mathcal{S}_{BS}\!= \frac{D}{\pi}v\!\int_{0}^{\beta}\!\!\!d\tau \cos[2\phi_{3}(0,\tau)].
\end{eqnarray}


{\it Effective model.---} Next, let us implement all the reasonable assumptions in order to derive an effective model (see Supplemental Material for details~\cite{suplement}). 
First, we present $\phi_{1,i}(\phi_{2,i})$ in terms of the new fields $\phi_{1,c},\phi_{1,s},..., \phi_{1,f}$ ($\phi_{2,c},\phi_{2,s},..., \phi_{2,f}$), which characterize the charge, spin, and other flavor modes. 
Interestingly, only the charge $\phi_{\alpha,c}=(N_{\alpha})^{-1/2}\sum_{i=1}^{N_{\alpha}}\phi_{\alpha,i}$ ($\alpha=1,2$) and $\phi_{3}$ modes enter the Coulomb blockade action, while all other flavor modes are excluded from further consideration. The coefficients $\sqrt{N_{\alpha}}$, which appears in Eq.~(\ref{aC}) after the redefinition of variables, effectively renormalize the Luttinger interaction parameters of the charge fields $\phi_{1,c}$, $\phi_{2,c}$: $K=\nu \rightarrow \nu N_{\alpha}$ (one can see it by rescaling $\sqrt{N_{\alpha}}\phi_{\alpha,c}\rightarrow \phi_{\alpha,c}$). 

Since the action $\mathcal{S}$ is Gaussian everywhere except for the $x=0$ point, a further simplification of the problem to one that is local in space but not local in time is possible by integrating out all the degrees of freedom $\phi_{1,c}(x\neq 0,\tau),\phi_{2,c}(x\neq 0,\tau),\phi_{3}(x\neq 0,\tau)$. 
After this procedure, the resulting action $\mathcal{S}$ can be diagonalized by a unitary transformation, such that it reduces $(\phi_{1,c},\phi_3,\phi_{2,c})$ to $(\phi_A,\phi_B,\phi_C)$~\cite{suplement}. 

After the rescaling $\phi_{\eta}/\sqrt{\nu}\rightarrow \phi_{\eta}$ ($\eta=A,B,C$), the effective actions turn into
\begin{eqnarray}\label{actioneff1}
&&\mathcal{S}_0 =\sum_{\eta=A,B,C}\sum_{\omega_n} \frac{|\omega_n|T}{\pi}\phi_{\eta}(-i\omega_n)\phi_{\eta}(i\omega_n),\\
&&\mathcal{S}_{C} = \frac{\nu E_{C}}{\pi^2}\int_{0}^{\beta}d\tau\left\{A_-\phi_{B}^{2}(\tau)+A_+\phi_{C}^{2}(\tau)\right\},\\
&&\mathcal{S}_{BS}\!= \frac{Dv}{\pi}\!\int_{0}^{\beta}\!\!\!d\tau \cos\left\{\sqrt{4\nu}\left[A_A\phi_A-A_B\phi_B-A_C\phi_C\right]\right\},\label{actioneff3}\nonumber\\
\end{eqnarray}
where $\phi_{\eta}(i\omega_n)=\int_0^{\beta}d\tau \phi_{\eta}(\tau)\exp(i\omega_n\tau)$ and $\omega_n=2\pi T n$ is the Matsubara frequency; $A_{\pm}=1+(N_1+N_2)/2\pm\sqrt{1+(N_1-N_2)^2/4}$, and coefficients $A_{\eta}$ are explicitly presented in \cite{footnote}. 
Thus, we found the effective action for three modes, two of which are gapped due to the Coulomb blockade.


{\it Strong interdot coupling.---} Let us estimate the linear conductance through the 2SCK circuit in the zero-frequency limit. 
Using the current conservation (Kirchoff's) law, the Kubo formula can be presented
\begin{eqnarray}\label{kubo}
G=G_{\rm max}\frac{2T}{\pi i}\lim_{\omega\rightarrow0}\omega \lim_{i\omega_n\rightarrow \omega +i0^{+}}\langle \phi_A(i\omega)\phi_A(-i\omega)\rangle,
\end{eqnarray}
where $G_{\rm max}=\nu G_0 N_1N_2/(N_1+N_2+N_1N_2)$, and $G_0=R_0^{-1}=e^2/2\pi$ is a unitary conductance. 
We treat Eq.~(\ref{kubo}) by using the functional integration technique and by utilizing the perturbation theory over the weak backscattering amplitude, $v\ll1$. 
Applying Wick's theorem and performing the Gaussian integration of all the fields,  in the low-temperature limit $T\ll E_C$ we obtain~\cite{suplement}
\begin{eqnarray}\label{res1}
G(T)=G_{0}K-G_{0}K^2\tilde{v}^2
\mathcal{C}_{\rm sc}\frac{\sqrt{\pi}\Gamma(K)}{2\Gamma\left(K+\frac{1}{2}\right)}\left(\frac{\pi T}{\nu E_C}\right)^{2K-2}.
\end{eqnarray}
Here, $K=\nu N_1N_2/(N_1+N_2+N_1N_2)$ is an effective Luttinger parameter, $\tilde{v}=v(\nu E_C/D)^{\nu-1}$ is the interaction-renormalized backscattering amplitude (the Kane-Fisher effect~\cite{kanefisher,kanefisheradd}), $\Gamma(x)$ is a gamma-function, and 
$\mathcal{C}_{\rm sc}(\nu,N_1,N_2)=\left(\gamma /\pi\right)^{\frac{2\nu(N_1+N_2)}{N_1+N_2+N_1N_2}}\left(A_+^{A_C^2}A_-^{A_B^2}\right)^{2\nu}$, where $\gamma=\exp(0.577)$ is the Euler's constant. 

We can compare formula~\eqref{res1} with the existing results by noting, that Eq.~(\ref{res1}) coincides with the expression describing the conductance of a multichannel single-site CK circuit in the FQH regime~\cite{parafilo4} when one of the QDs connects to an infinite number of channels ($N_1\rightarrow \infty$). Indeed, in this case $K\rightarrow \nu N_2/(1+N_2)$ and $A_-=A_B^2=1+N_2$, while $A_C^2\rightarrow0$ \cite{footnote}. 
In the other limiting case $N_1=N_2=1$ and in the IQH regime $\nu=1$, Eq.~(\ref{res1}) coincides with the first two terms in Eq.~(D7) derived in~\cite{karki1} and with Eq.~(3b) reported in \cite{karki2}.

Moreover, Eq.~\eqref{res1} resembles the conductance in the LL with a weak potential barrier, when the interaction parameter $K$ is determined by the fractional filling factor $\nu=1/m$ and by the number of ballistic channels that enter both QDs, $N_1$ and $N_2$. This can be attributed to a well-known result by Safi and Saleur~\cite{safisaleur}. 
The Luttinger interaction parameter $K'$ will experience renormalization if the quantum conducting channel formed in the LL is set in series with the resistance $R_{\rm tot}$ (dissipative environment) as $K=1/(K'^{-1}+R_{\rm tot})$. Indeed, bare interaction in the system is $K'=\nu$, while $N_1$~($N_2$) channels attached to the first (second) QD can be treated as $N_1$ ($N_2$) resistors, each with the resistance $r=(\nu G_0)^{-1}/R_0$, coupled in parallel, $r_{\alpha}=1/(\sum_{i=1}^{N_{\alpha}}r^{-1})$ [see Fig.\ref{Fig1}(b)]. 
As a result, all the open channels implement the total resistance $R_{\rm tot}=r_1+r_2$, and thus, $K=1/[\nu^{-1}+(\nu N_1)^{-1}+(\nu N_2)^{-1}]$. 

The perturbation theory justifies Eq.~(\ref{res1}) down to the `boundary' temperature $T_B \sim \nu E_C \tilde{v}^{1/(1-K)}$, while descending down to even lower temperatures requires the consideration of the model with a tunnel barrier between two QDs.


\begin{figure}
\centering 
\includegraphics[width=0.9\columnwidth]{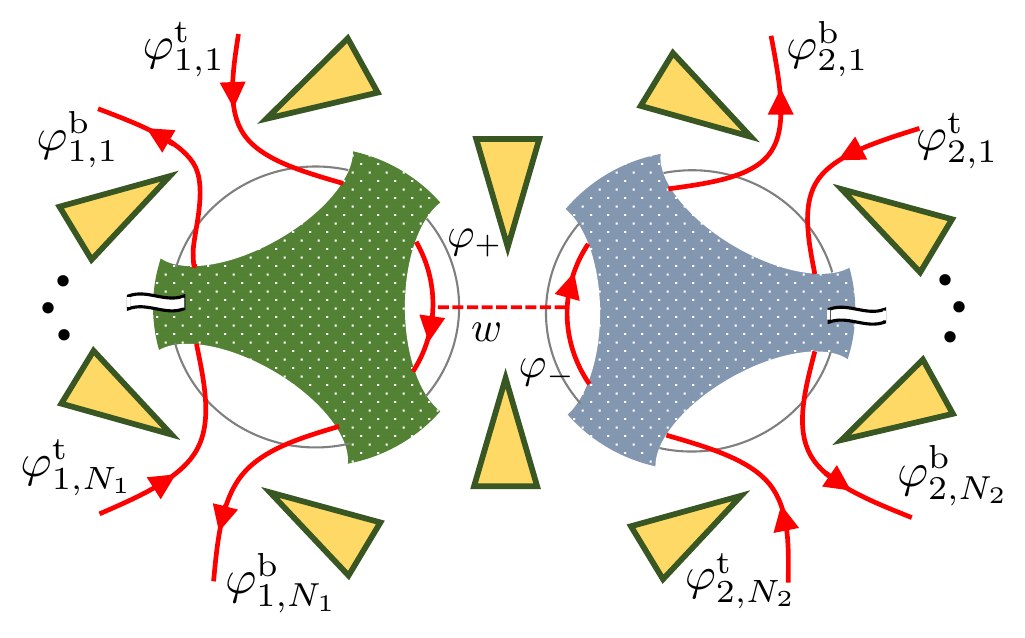} \caption { The device presented in Fig.\ref{Fig1}(a) in the case of tunnel coupling between two QDs.}
\label{Fig2} 
\end{figure}

{\it Weak interdot coupling.---} Let us consider the temperatures  $T\ll T_B$, when the coupling between two QDs is due to tunneling, $w=\sqrt{1-v^2}\ll 1$. In this regime, it is more convenient to calculate the conductance using the Hamiltonian formalism. 
The total Hamiltonian of the system shown in Fig.~\ref{Fig2} reads $\tilde{H}=\tilde{H}_0+\tilde{H}_C+\tilde{H}_{\rm tun}$, where the first term is a chiral LL Hamiltonian of the $2(N_{1}+N_{2}+2)$ edge modes,
\begin{eqnarray}
&&\tilde{H}_0=\frac{v_F}{4\pi \nu}\sum_{\alpha=1,2}\sum_{i=1}^{N_{\alpha}+1}\int dx \left\{\left(\partial_x \varphi_{\alpha,i}^{\textrm{t}}\right)^2+\left(\partial_x \varphi_{\alpha,i}^{\textrm{b}}\right)^2\right\},\nonumber\\
\end{eqnarray}
while the second term is the charging energy of 2SCK circuit,
\begin{eqnarray}
\tilde{H}_C=\frac{E_C}{4\pi^2}\left(\hat{n}_1^2+\hat{ n}_2^2\right)\,,\,
\hat{n}_{\alpha}=\sum_{i=1}^{N_{\alpha}+1}\left[\varphi_{\alpha,i}^{\rm t}(0)-\varphi_{\alpha,i}^{\rm b}(0)\right].\nonumber\\
\end{eqnarray}
Since the QPC between two QDs pinches off the 2SCK circuit into two weakly coupled subsystems, four bosonic fields $\varphi_{N_{\alpha}+1}^{\rm t(b)}$ describe two `looped' edge states. 
The term `looped' means that the outgoing edge mode, say $\varphi_{N_1+1}^{\rm t}$, from the first QD after nearly perfect reflection at the QPC enters back the first QD as an incoming state, $\varphi_{N_1+1}^{\rm b}$. Thus, the fields $\varphi_{N_{\alpha}+1}^{\rm t}$ and $\varphi_{N_{\alpha}+1}^{\rm b}$ can only differ by an insignificant phase factor \cite{florian}. Both these fields merge into one looped field, which we denote as $\varphi_{+}$ ($\varphi_-$ is used for the looped state of the second QD) in what follows, see Fig.~\ref{Fig2}. 
Nevertheless, these two looped states participate in the tunneling events and `contain the information' about other modes that enter the QDs. 
Thus, the single-electron tunneling between two QDs can be described by the Hamiltonian
\begin{eqnarray}
\tilde{H}_{\rm tun}=\frac{D}{2\pi v_F}w\left\{e^{i\nu^{-1}[\varphi_{+}(0)-\varphi_-(0)]}+\textrm{H.c.}\right\}.
\end{eqnarray}

The average tunneling current in the lowest-order of perturbation theory reads
\begin{eqnarray}\label{currdef}
I=\frac{e |w|^2D^2}{(2\pi v_F)^2}\int dt \left\{e^{ieVt}\mathcal{G}_+(t)\mathcal{G}^{\dag}_-(t)-(t\rightarrow-t)\right\},
\end{eqnarray}
where $V$ is a bias voltage between two QDs, and the correlation function $\mathcal{G}_{\pm}(t)$ is given by (see, e.g., \cite{florian,arthur,eugene})
\begin{eqnarray}\label{corr}
&&\mathcal{G}_{\pm}(t)=\langle e^{i\varphi_{\pm}(0,t)/\nu}e^{-i\varphi_{\pm}(0,0)/\nu}\rangle = \left(\frac{\pi T}{D}\right)^{\frac{1}{\nu}}\\ &&\times\left(\frac{\pi^2 T }{\gamma N_{1(2)}\nu E_C}\right)^{\frac{2}{\nu N_{1(2)}}}
\left[\frac{1}{i\sinh[\pi T(t-i0^+)]}\right]^{\frac{1}{\nu}+\frac{2}{\nu N_{1(2)}}}.\nonumber
\end{eqnarray}
%

After substituting Eq.~(\ref{corr}) in Eq.~(\ref{currdef}), a straightforward calculation gives~\cite{suplement}
\begin{eqnarray}\label{res3}
G=G_{\rm tun}\mathcal{C}_{\rm wc}\frac{\sqrt{\pi}\Gamma\left(\frac{1}{K}\right)}{2\Gamma\left(\frac{1}{2}+\frac{1}{K}\right)}\left(\frac{\pi T}{\nu E_C}\right)^{2/K-2},
\end{eqnarray}
where $G_{\rm tun}=e^2 |\tilde{w}|^2/(2\pi v_F^2)$, with $\tilde{w}=w(\nu E_C/D)^{1/\nu-1}$ a renormalized tunnel amplitude in accordance with the Kane-Fisher effect \cite{kanefisher,kanefisheradd}, and $ \mathcal{C}_{\rm wc}(\nu, N_1,N_2)=\left(\pi/\gamma\right)^{\frac{2}{K}\frac{N_1+N_2}{N_1+N_2+N_1N_2}}N_1^{-2/\nu N_1}N_2^{-2/\nu N_2}$. 

Formula~(\ref{res3}) is consistent with the previous studies \cite{thanh2018,parafilo4,kiselev}. 
In particular, if $N_1\rightarrow \infty$, Eq.~(\ref{res3}) recalls Eq.~(21) in \cite{parafilo4}. In the IQH regime ($\nu=1$), Eq.~(\ref{res3}) scales as $G(T)\propto T^{2(N_1+N_2)/N_1N_2}$ -- the scaling, which was for the first time predicted in~Ref.~\onlinecite{thanh2018} for the cases $N_1,N_2=1,2$, and in \cite{parafilo4, kiselev} for the arbitrary number of conducting channels coupled to both QDs. In addition, Eqs.~(\ref{res1}) and (\ref{res3}) at the limiting case $N_1$$=$$N_2$$=$$1$ ($\nu=1$) are in agreement with the universal crossover scaling of the conductance calculated by the Bethe ansatz in \cite{karki1} [see Eq.~(27) there] and \cite{fendleyprl,fendley}.

The similarity of Eq.~(\ref{res3}) with the backscattering correction in Eq.~(\ref{res1}) is simply the $K$ to $K^{-1}$ correspondence, well-known for the LL with `weak' and `strong' potential barriers \cite{giamarchi,safisaleur}. 
Both the formulas~\eqref{res1} and~\eqref{res3} cover the temperature dependence of the conductance through the 2SCK circuit at two temperature regimes, $T\gg T_B$ and $T\ll T_B$.

It is rather easy to explain the power-law temperature dependence of Eqs.~(\ref{res1}) and~(\ref{res3}), and their relation to the LL physics. 
Indeed, the fluctuations of two gapped modes $\phi_B(0,\tau)$ and $\phi_C(0,\tau)$ in the effective model given by Eqs.~(\ref{actioneff1})--(\ref{actioneff3}) are suppressed due to the Coulomb blockade. 
At low temperatures, $T\ll E_C$, these degrees of freedom can be integrated out. 
The reduced model (see Eq.~(23) in~\cite{suplement}), similar to the impurity problem in LL with Luttinger parameter $K$~\cite{kanefisher}, is referred to as the boundary sine-Gordon (BSG) model in the literature~\cite{gogolin}. 
The BSG model describes `similar physics' in a broad class of physical systems (see, e.g.,~\cite{chakra,shmidt,grabert1,QBM}),
including a quantum impurity embedded in a 1D nanowire~\cite{giamarchi} or single conducting channel coupled to an Ohmic dissipative environment~\cite{safisaleur,weiss,devoret}.
The comparison between different systems, defined by the BSG model, makes it possible to simulate~\cite{pierre,DCB}, interpret~{\cite{safisaleur,parafilo4}}, and even predict physical effects.


Interestingly, the 2SCK device can also be treated as a mesoscopic circuit simulating the tunneling between two different LLs equivalent to the tunneling between two FQH edges with different filling factors~\cite{chamon}. Each LL is determined by $N_{\alpha}$ channels connected to $\alpha$th QD [the corresponding interaction constant reads $K_{\alpha}=\nu N_{\alpha}/(2+N_{\alpha})$]. In this case, $K$ from Eqs.~(\ref{res1}), (\ref{res3}) can be presented as the reduced Luttinger parameter, $K=2/(K_1^{-1}+K^{-1}_2)$ in accordance with \cite{chamon}. Thus, Eqs.~(\ref{res1}), (\ref{res3}) can be used to analyze transport properties of two LLs coupled via weak or strong potential barriers in the spirit of experiments performed in \cite{pierre,DCB}.

Concluding this subsection, let us pose an important question: Is it possible to account for the effects of interaction at the quantum critical point of the 2SCK problem when the non-Fermi liquid behavior prevails? 
It can only be done in the particular case of $N_1$$=$$N_2$$=$$1$ channels when the system is tuned to the so-called {\it triple point} \cite{karki1,karki2}. 
As it was shown~\cite{karki1,karki2}, this critical point corresponds to the existence of the fractionalized ($\mathbb{Z}_3$ parafermion) excitation, while the whole problem can be exactly solved by the analogy of Emery-Kivelson solution \cite{emerykivelson} in the Toulouse limit.

{\it Triple quantum critical point.---} Let us consider the situation when the 2SCK circuit operates in the FQH regime with $\nu=1/m$, while $N_1$$=$$N_2$$=$$1$. We account for the gate voltage dependence of the charging energy by considering the weak backscattering at three QPCs, $v_{\alpha}\ll1$ ($\alpha=1,2,3$). For this, one needs to replace 
$\hat{n}_{\alpha}\rightarrow \hat{n}_{\alpha}-\pi \mathcal{N}_{\alpha,g}$ in Eq.~(\ref{aC}), where $\mathcal{N}_{\alpha,g}$ is the dimensionless gate voltage parameter at $\alpha=1,2$ QD. The half-integer value $\mathcal{N}_{\alpha,g}=1/2+n$ ($n$ is integer) corresponds to the charge degeneracy between $n$ and $n+1$ charge states in $\alpha$th QD. We also use $\mathcal{S}'_{BS}=(D/\pi)\sum_{\alpha=1}^{3}\int d\tau v_{\alpha}\cos[2\phi_{\alpha}(0,\tau)]$ instead of Eq.~(\ref{backsc}) (here, $\phi_{\alpha,1}\equiv \phi_{\alpha}$ with $\alpha=1,2,3$). 

Following the calculations made in Ref.~\onlinecite{karki1}, we find 
\begin{eqnarray}\label{res2}
G=\frac{\nu G_0}{3}\left\{1-\mathcal{C}_1\left(\frac{\pi T}{\nu E_C}\right)^{\frac{2\nu}{3}-2}
-\mathcal{C}_2\left(\frac{\pi T}{\nu E_C}\right)^{\frac{2\nu}{3}}\right\},\nonumber\\
\end{eqnarray}
where $\mathcal{C}_{1(2)}=(\nu/3)\,3^{\nu/3}(\gamma/\pi)^{4\nu/3}\sqrt{\pi}\Gamma(\nu/3)/[2\Gamma(\nu/3+1/2)]\cdot\tilde{\mathcal{C}}_{1(2)}$ (constants $\tilde{\mathcal{C}}_1$ and $\tilde{\mathcal{C}}_2$ are presented in \cite{suplement} and \cite{footnote2}) are $\nu$ and $\mathcal{N}_{\alpha,g}$-dependent constants fully consistent with Eqs.~(D8) and~(D9) in Ref.~\onlinecite{karki1}. Thus, Eq.~(\ref{res2}) generalizes the universal conductance scaling in the vicinity of a quantum critical (triple) point in the case of the FQH regime. The conductance scaling of the first two terms in Eq.~(\ref{res2}) coincide with Eq.~(\ref{res1}), since $K\rightarrow \nu /3$ at $N_1=N_2=1$. At $\nu=1$, the conductance Eq.~(\ref{res2}) reduces to the scaling Eq.~(13) from~\cite{karki2}.

At the tri-critical point, $\mathcal{N}_{1,g}=\mathcal{N}_{2,g}=1/2$, $|v_1|=|v_2|$ and $v_3=2\cdot 3^{-\nu/2}|v_1|$,
(here, we choose the simplest set of parameters; the detailed analysis of the triple point for other sets of parameters can be found elsewhere~\cite{karki1}), 
the constant $\mathcal{C}_1$ vanishes, while the third term in Eq.~(\ref{res2}) associated with the least irrelevant perturbation at the quantum critical point survives. 
Thus, Eq.~(\ref{res2}) at the triple point can be treated as a universal Kondo scaling of the 2SCK model in the FQH regime, $G(T)=G_{\rm max}\{1-(T/T_K)^{2\nu/3}\}$, with the Kondo temperature $T_K\propto \nu E_C /|\tilde{v}_1|^{3m}$ ($m$ is odd integer, which comes from $\nu=1/m$). 
At $T\gg T_K$, we would expect a logarithmic correction $\log^{-2}(T/T_K)$ to the conductance. Far from the tri-critical point, the second term in Eq.~(\ref{res2}) prevails over the third one. In this case, the Kondo screening is suppressed either by channel asymmetry ($|v_1|\neq|v_2|\neq|v_3|$) or by pseudomagnetic field (gate voltages $\mathcal{N}_{1,g}$ and $\mathcal{N}_{2,g}$ that lift charge degeneracy in the QDs). 
An exaggerated channel asymmetry ($|v_3|=v$, $|v_1|=|v_2|=0$) allows us to formulate a universal conductance scaling, see Eqs.~(\ref{res1}) and~(\ref{res3}), in the case $N_1\neq1, N_2\neq1$.



{\it Conclusions.} We studied the effects of interaction in a 2SCK simulator by considering a hybrid metal-semiconductor double-dot setup in the FQH regime with a filling factor $\nu=1/m$ ($m$ is an odd integer). 
We showed that the number of open ballistic channels $N_1,N_2$ connected to both the QDs uniquely determines the transport properties at low temperatures. The conductance scaling at the strong $KG_0-G(T)\propto T^{2K-2}$ and weak $G(T)\propto T^{2/K-2}$ inter-dot coupling is fully featured by the effective interaction parameter $K=\nu N_1N_2/(N_1+N_2+N_1N_2)$. 
Meanwhile, the 2SCK circuit can be treated as a simulator of tunneling between two different LLs, whose interaction parameters are defined by $N_1$ and $N_2$, respectively. 
Moreover, in a particular case of $N_1=N_2=1$ and $\nu=1/m\neq1$, we predict how the finite temperature correction at the triple quantum critical point, characterized by the emergence of fractional $\mathbb{Z}_3$ excitation, is determined by the fractional filling factor $\nu$: $1-G(T)/G_{\rm max}\sim (T/T_K)^{2\nu/3}$.

{\it Note added.} -- Ref.~\cite{kiselev} addresses a closely related problem of transport properties in a multichannel two-site charge Kondo circuit in the case of weak interdot coupling.



\textit{Acknowledgements.} The authors would like to thank Jaeho~Han, A.~Andreanov, Dung~X.~Nguyen and F.~St\"abler for fruitful and inspiring discussions. A.V.P. acknowledges support by the Institute for Basic Science in Korea (IBS-R024-D1). 
\textcolor{black}{V.M.K. is grateful to the Project FSUN-2023-0006
and ``BASIS'' Foundation.}


\begin{thebibliography}{10}
\bibitem{footnote} 
Coefficients $A_{\eta}$ ($\eta=A,B,C$) in $\mathcal{S}_{BS}$ explicitly read as $A_A=\sqrt{N_1N_2/(N_1+N_2+N_1N_2)}$ and $A_{B(C)}=(1+A\mp\Delta)/[N_2\sqrt{N_2+(1+A\mp\Delta)^2+N_1(A\mp\Delta)^2}]$, where we denote $A=(N_1-N_2)/2$ and $\Delta=\sqrt{1+A^2}$.
\bibitem{footnote2}
Here, we present explicit expressions for constants $\tilde{\mathcal{C}}_{1(2)}$: $\tilde{\mathcal{C}}_1(\nu,\tilde{v}_{\alpha}, \mathcal{N}_{i,g})=\,\{\tilde{v}_1^2+\tilde{v}_2^2+3^{\nu}\,\tilde{v}_3^2+2\,\tilde{v}_1\tilde{v}_2\cos[2\pi(\mathcal{N}_{1,g}+\mathcal{N}_{2,g})]+2\cdot3^{\nu/2}\tilde{v}_3\,\left(\tilde{v}_1\cos[2\pi \mathcal{N}_{1,g}]+\tilde{v}_2\cos[2\pi \mathcal{N}_{2,g}]\right)\}$,
and $\tilde{\mathcal{C}}_2(\nu,\tilde{v}_{\alpha}, \mathcal{N}_{i,g})=(\nu \pi^2/27)\,[2\nu/(2\nu+3)]\{28(\tilde{v}_1^2+\tilde{v}_2^2)+4\cdot 3^{\nu}\,\tilde{v}_3^2-52\,\tilde{v}_1\tilde{v}_2\cos[2\pi(\mathcal{N}_{1,g}+\mathcal{N}_{2,g})]-8\cdot3^{\nu/2}\tilde{v}_3\,\left(\tilde{v}_1\cos[2\pi \mathcal{N}_{1,g}]+\tilde{v}_2\cos[2\pi \mathcal{N}_{2,g}]\right)\}$. See also Ref.~\cite{suplement}.
\end{thebibliography}
\end{document}